# A self-sensing microwire/epoxy composite optimized by dual interfaces and periodical structural integrity

Y. J. Zhao[*], X. F. Zheng[*], F. X. Qin[†], D. Estevez, Y. Luo, H. Wang,

and H. X. Peng[‡]

*Institute for Composites Science Innovation (InCSI), School of Materials Science and Engineering, Zhejiang University, Zheda Road, Hangzhou, 310027, PR. China*

## Abstract

Self-sensing composites' performance largely relies on the sensing fillers' property and interface. Our previous work demonstrates that the microwires can enable self-sensing composites but with limited damage detection capabilities. Here, we propose an optimization strategy capitalizing on dual interfaces formed between glass-coat and metallic core (inner interface) and epoxy matrix (outer interface), which can be decoupled to serve different purposes when experiencing stress; outer interfacial modification is successfully applied with inner interface condition preserved to maintain the crucial circular domain structure for better sensitivity. We found out that the damage detection capability is prescribed by periodical structural integrity parameterized by cracks number and location in the case of damaged wires; it can also be optimized by stress transfer efficiency with silane treated interface in the case of damaged matrix. The proposed self-sensing composites enabled by a properly conditioned dual-interfaces are promising for real-time monitoring in restricted and safety-critical environments.

---

[*] Equally contributing authors
[†] Corr author: faxiangqin@zju.edu.cn
[‡] Corr author: hxpengwork@zju.edu.cn

## 1. Introduction

Composites reinforced by carbon fibers and glass fibers have been widely used in aerospace industry because of their low density and superior mechanical performance over metallic materials[1]. Moreover, the proportion of fiber-reinforced composites in aircraft has exceeded 50%, which could lead to catastrophic consequences if damages appear and extend in the composites, degrading their structural integrity and worsening their mechanical performance. Thus, a life-safety effective and economical way of real-time inspection is needed for fiber-reinforced composites, commonly referred as structural health monitoring (SHM). Currently widely-used SHM technologies are mainly based on embedded sensors like optical fibers, piezoelectric films, magnetostrictive materials and memory alloys [2-5]. Due to extended functionalities enabled by these embedded sensors, their composite materials have obtained myriad designations as 'self-sensing' and 'smart', which can be used particularly in non-destructive monitoring of local stresses. However, such inclusions have many setbacks in practical utilization, including poor mechanical properties of the composites caused by the size mismatch between the sensors (usually no less than 100 microns) and carbon fibers or glass fibers (several microns), complicated and pricy signal processing and conditioning, limited service environments caused by the brittleness of optical fibers as well as easy polarization and easy electrical breakdown of piezoelectric films. As such, alternative effective techniques are demanded both economically and technologically, with the least harm to the integrity and mechanical performance of the composites [6-8].

In recent years, much research effort has been dedicated to the investigation of engineered composites containing ferromagnetic glass-coated microwires in terms of their tailorable size (2 μm ~100 μm), sensitivity to external stimuli such as, magnetic field, stress or temperature and microwave tunability[9-11]. The magnetic field and stress-dependent high frequency impedance properties of such wires make them promising candidates for SHM applications [12-14]. Two monitoring methods have been developed based on such effects, "wired" and "wireless", in which the impedance of the microwire is measured through electrical contacts and through the microwire scattering properties in free space, respectively [15-17]. The latter method offers a more efficient, reliable and accurate remote monitoring technique owing to the absence of cable connections and network analyzer correction issues. However, a better understanding of the correlation between specific damage attributes (size, distribution, etc.) and electromagnetic parameters required to assess the structural integrity of the composite is still missing. The difficulties in mapping the damage through the electromagnetic signature are largely related to heterogeneity of properties between the matrix and the microwire, and improper stress transfer associated with microwire-matrix interfacial behavior. Thus, a design strategy integrating the properties of microwire and composite mechanical behavior needs to be adopted. In this work, to efficiently enhance the damage and local structural integrity sensing capabilities of microwire composites, an approach based on dual interfacial optimization is proposed to maintain, if not improve, the magnetic features of wires via the inner interface and better the outer interface bonding. The damage was introduced in a controlled way and evaluated through remote microwave interrogation, first on free-standing microwire arrays with preset cracks and then on epoxy matrix composites incorporating silane-modified glass-coated microwire arrays. Thanks to the existence of glass coat that served as a sensitivity supplier for free-standing microwire and a mediator for matrix-wire interactions, the structural integrity sensitivity of the microwire composites increased substantially while preserving both the wire magnetic properties responsible for sensing capabilities and mechanical integrity of the composites.

## 2. Experimental details

### 2.1. Design of free-standing microwire arrays with preset cracks

To prepare the sample of free-standing microwire arrays for the rectangular waveguide WR90 (0.9 in× 0.4 in), a special kraft paper frame was designed in which a rectangular window was cut (22.86 mm×10.16 mm) at the center, matching the size of the waveguide frame, vertical grids were also drawn to ensure the wires were properly placed. (错误!未找到引用源。a). Five parallel continuous $Co_{68.7}Fe_4Si_{11}B_{13}Ni_1Mo_{2.3}$ glass-coated microwires, whose inner and total diameters are 19.3 μm and 23.2 μm, respectively, were fixed along the lines on the paper with a 2 mm spacing among them. Then, preset cracks were introduced by cutting the microwires along their length at a specific position to simulate the damage locating right on the microwires. The microwires were labeled A to E from left to right, respectively. The number following the capital letter represents the distance in mm between the crack and the top end of the microwire inside the waveguide frame, e.g. A3 (crack in wire A, 3 mm from the top), A3B3 (crack on wire A, 3 mm from the top and crack on wire B, 3 mm from the top) (Fig. 1b). The whole series of crack locations on the microwire arrays is summarized in 错误!未找到引用源。.

### 2.2 Preparation of surface-modified microwires and pull-out samples

The glass-coated amorphous wires were treated with silane coupling agent to study the interfacial bonding with the epoxy matrix and therefore optimize the stress transfer in the composites. To prepare the silane solution for the surface modification of the glass-coated microwire, KH550 silane was diluted with ethanol to 2 wt.% and 5 wt.% (错误!未找到引用源。 and Fig. 1c). Then, the amorphous microwires were immersed in the silane solution and sonicated for 20 minutes. To prepare the pull-out samples for microwire/epoxy bonding strength analysis, a grid paper with a rectangular window (10 mm × 10 mm) was made (错误!未找到引用源。d). A 10 mm long microwire was embedded into commercial LT-5028 epoxy resin (100:30 epoxy/hardener) at one end and stuck by cyanoacrylate adhesive to the paper at the other end, ensuring that the effective pull-out length of the microwire was 1 mm. After that, all the pull-out samples were cured at 50 °C for 4 h and 700°C for 6 h. Then, tabs made of kraft paper were stuck to the both ends of the wire inside the grip section to guarantee a uniform load distribution.

## 2.3 Preparation of smart self-sensing composites containing microwire arrays

To prepare the composites which can be drilled easily and accurately as well as locate the damage near the microwire, commercial LT-5028 epoxy resin was selected as the matrix of the composites. Four types of composite samples were prepared. In the reference sample (matrix), the epoxy resin was poured into a designed cuboid mold to produce samples with dimensions of 22.86 mm (length) × 10.16 mm (width) × 2 mm (thickness). The rest of the samples were prepared in the same manner except that five parallel microwires were buried along the width and at the center of thickness direction with a wire spacing of 2 mm. As cast microwires, and wires modified by 2 wt.% and 5 wt.% silane solution, were incorporated in the matrix, respectively. All the samples were then cured at 50 °C for 4 h and 70 °C for 6 h. After that, the drilled samples were prepared by an electric drill with 1 mm diameter bit. Two drilled composites were studied. The distances between the center of the hole and the nearest microwire were 2 mm and 1 mm respectively, as shown in Fig. 1e.

[Fig.1 here]

## 2.4 Characterization techniques

The morphology of the pull-out microwires was investigated by scanning electron microscopy (SEM), Hitachi 3400, Japan. Pull-out tests for surface-modified microwires were performed with an

INSTRON tensometer 5943 at a travel rate of 1 mm/min at room temperature. The surface characterization of the samples was performed by X-ray photoelectron spectroscopy (Kratos AXIS Supra) and contact angle measurements (DataPhysics Instruments OCA 20). The scattering parameters of all samples were measured by the Rohde&Schwarz ZNB 20 vector network analyzer (VNA) using WR90 waveguide in TE 10 dominant mode frequency range [18, 19] from 8.5 GHz to 11 GHz. The VNA was calibrated by the thru-reflect-line (TRL) [20] method and the measurements were performed on transmission mode. The power reflection and absorption coefficients can be directly computed from the scattering parameters as R = $|S_{11}|^2$ and A = 1 − $|S_{11}|^2$ − $|S_{21}|^2$, respectively.

[table 1] and [table 2] here

## 3 Results and discussions

### 3.1 Microwave properties of the free-standing microwire arrays with preset cracks

To investigate the effects of damage which locates right on the microwires, various crack distributions were introduced into the free-standing microwire arrays. Considering the symmetry of the array consisting of five parallel microwires, here we only discuss the microwave performance of the microwire arrays with one or two cracks to simulate the condition in which the microwire array composite still remains partial structural integrity. The crack distribution dependence can be evaluated both by using the absorption (Fig. 2a) and reflection (Fig. 2b) coefficients in all samples.

When only one crack was introduced into the array, the crack was preset either on microwire-A or on microwire-C of the array (defined as the x coordinate, from left to right), which divided the microwire to short parts with different lengths (defined as the y coordinate, from top to bottom). When both the x and y coordinates of the crack are determined, the location of the crack can be easily traced. Fig. 2a shows the absorption spectra of free-standing microwire arrays containing one crack as well as the uncut microwire arrays. For the uncut sample, which microwire length is 10.16 mm, the characteristic absorption peak should occur at around 9.3 GHz, corresponding to the Lorentz-type dielectric resonance [21-23]. When one crack is introduced to the microwire, it was split to two parts, and the longer part in the array system contributes to the dominant dielectric response due to a higher induced dipole moment. Therefore, the effective length for the samples A3 and C3 can be estimated as 7.16 mm while the length of A5 and C5 can be estimated as 5.16 mm. For all cut-samples, there are dual absorption peaks, the peak occurring at around 9 GHz may be caused by the presence of the paper frame which introduces a thin non-conductive gap between the two parts of the waveguide, regardless of the microwire length. Meanwhile, as the effective length decreases, the absorption dispersion becomes broader and the characteristic frequency of the second peak blueshifts to 10.1 GHz and 10.3 GHz for A3 and A5, 10 GHz and 10.3 GHz for C3 and C5, respectively, which corresponds to the characteristic frequencies determined by the antenna resonances [23]

$$f_{res,n} = \frac{c(2n-1)}{2l\sqrt{\varepsilon}}$$

where $c$ represents the light speed, $n$ natural number (for the lowest resonance n = 1), $l$ microwire length, and $\varepsilon$ the relative permittivity of the matrix. It is obvious that for microwire arrays with one crack, the characteristic frequency of absorption peak is determined by microwire length regardless of which microwire the crack locates at. The mismatch characteristic frequencies of A3 and C3 may be caused by the different tilted stages of the free ending in the microwire cuts.

Since the crack position along the wire can be deduced from the characteristic frequency of second peak of the absorption spectra, it is equally important to assess the x coordinate of the crack. It is found in the Fig. 2b that when only one crack locates on different microwires, the value of reflection coefficient varies due to different crack distributions (different x coordinate) despite the same effective microwire length (the same y coordinate). Compared to the uncut ferromagnetic microwire arrays, which owns the largest reflection coefficient over the 8.5-11 GHz frequency range due to the better periodicity of the intact microwires, the reflection value of array with a crack on microwire-A is lower than that of array with a crack on microwire-C. This is because the crack on microwire-C causes a larger spacing between microwire-B and microwire-D, making the original array into two smaller identical arrays (AB and DE). In this condition, the microwave is originally reflected and interfered by array-AB and array-DE, contributing to a larger reflection coefficient. On the contrary, despite a broken microwire, the periodical arrangement of microwires in the array remains the same as the previous one in A3 and A5, yielding a similar reflection.

[Fig. 2 here]

Similarities can be seen when two cracks locate on different microwires (Fig. 3). When two cracks are located on microwire-A and microwire-B or microwire-B and microwire-D, the crack pattern is defined as pattern-AB and pattern-BD, respectively. For arrays with two broken microwires, the absorption peaks (Fig. 3a) appear at lower frequency as the effective length becomes larger, which is consistent with the resonance frequency shift [23]. The absorption amplitudes of both patterns exhibit similar values, which are higher than that of the uncut array. With reference to Fig.3a, as the crack number grows, the absorption peak becomes less pronounced due to the degradation of the array integrity.

Microwave reflection coefficient of the array with two cracks (Fig. 3b) mainly depends on the relative distribution of the damage, causing different array patterns and therefore leading to different reflection values. The reflection dispersion in descending order of magnitude, uncut array, A3B3, B5D5 and A5B5, indicates that reflection coefficient is proportional to the integrity (which in context refers to longer effective microwire length) and the periodicity (which refers to the array with stronger interference of the electromagnetic wave) of the array.

[Fig. 3 here]

To summarize, for free-standing microwire arrays, by evaluating the absorption dispersion, the number of the crack can be, to some degree, predicted. Then, the characteristic frequency of the absorption peak indicates that the effective microwire length and crack pattern can be associated with the reflection coefficient amplitude.

### 3.2 Investigation of interfacial bonding strength and microwave properties of the microwire-enabled smart self-sensing composite

In practical applications, the self-sensing microwire is supposed to be embedded in matrix, introducing the capacity of structure health monitoring without altering significantly the mechanical properties of the composites. The interfacial surface becomes crucial, deciding the adhesion quality between microwires and the matrix as well as the effectiveness of stress transfer. Negative magnetostriction microwires may become sensitive to the tensile stress with or without the presence of a weak DC bias magnetic field [24, 25]. Therefore, adequate microwire/matrix adhesion which results in efficient stress transfer and optimal microwire compositions are necessary in order to exploit the capability of microwire-polymer matrix for remote sensing of damage distribution. Here, we investigated the interfacial bonding strength by pull-out test and microwave performance of the epoxy composites incorporating microwire arrays modified by two silane coupling agent concentrations (2 wt.% and 5 wt.%).

### 3.2.1. Interfacial bonding strength of silane-modified microwires

Pull-out tests are introduced to assess the adhesion quality between the epoxy matrix and the surface-modified glass-coated microwire. A pull-out test of a glass-coated microwire embedded into a polymer matrix reveals two adhesion interfaces, i.e. a dual interface. The first one between the wire metal core and its glass coating, and the second one between the glass coating and the outer epoxy layer. According to Fig. 4, concentration of the silane has pronounced influence on pull-out results. Since both glass coating and metal core are brittle, the interface condition between glass coating/epoxy can be evaluated by plastic deformation. The load-displacement curve of the as-cast microwire appears generally linear with fluffy steps before fracture (Fig. 4), indicating that only a small part of glass coating broke during the pull-out process while the bonding between the glass coating and the metal core dominated. For microwires treated by 2 wt.% silane, the curve remains a linear trend but with a sudden drop of the load at around 0.17 mm of displacement, which can be attributed to the fracture of glass coating and sliding of the glass coating/metal core interface. When the silane concentration rises into 5 wt.%, the curve shows a linear trend associated to elastic deformation with a small shoulder just before fracture (zoomed view, inset of Fig. 4) due to the ductile fracture of the glass coating/epoxy interface. As the load exceeded the bonding strength of the glass coating/metal core interface, the metal core was pulled out completely showing the most enhanced glass coating/epoxy interfacial adhesion of all the samples which will ultimately provide the most effective epoxy/glass- coated microwire stress transfer. Among all the microwires, 2 wt.% silane modified sample bears the largest load, with the maximum at 1.2 N. In this case, the bonding strength of glass coating/epoxy interface overcomes the bonding strength of the glass coating/metal core interface, resulting in the largest pull-out strength compared to the amorphous wire and 5wt.% silane-modified sample, in which glass coating/metal core interface and metal core, respectively contribute to the pull-out loading capacity.

[Fig. 4 here]

The scanning electronic microscope (SEM) images (Fig. 5) of the pulled-out sample confirm the explanation given above to the load-displacement curves. Fig. 5a shows that the as-cast microwire has been totally pulled out, wrapped with both glass-coatings and rough epoxy lumps, which indicates poor interfacial adhesion and inefficient stress transfer. When the silane concentration is 2 wt.% (Fig.5b), the resultant smaller diameter implies that the pulled-out fiber is mostly metal core with some tiny cracks on the surface. When the bonding strength of both interfaces (epoxy/glass coating and glass coating/metal core) is comparable, the glass coating breached and slid between the epoxy layer and metal core, resulting in crack propagation along the effective pulled part. Most of the glass coating adhered to the epoxy, while some glass shards remained on the pulled-out metal core surface. As the concentration increases to 5 wt.% (Fig. 5c), the smaller diameter and smoother surface indicated that the pulled-out fiber was metal core without glass coating or epoxy.

[Fig. 5 here]

To further understand the mechanism of bond formation of silane to the glass-coated wire and thus its relationship to interfacial bonding strength, surface characterization by X-ray photoelectron spectroscopy (XPS) and contact angle (CA) were performed. In order to avoid distortion from the uneven surface of the microwire and thus obtain more accurate results, we replicated the same silane treatment conditions applied on the glass-coated microwire to a flat Pyrex glass substrate. As expected, the N1s XPS spectrum of the glass substrate without silane surface modification (Fig. 6a) does not show any significant signal in the spectral region. Primary and secondary amines, $-NH_2$ and $-NH$, are detected for the 2 wt.% silane modified glass substrate (Fig. 6b) at binding energies of 399.39 eV and 400.56 eV, respectively, indicating the presence of the silane coupling agent. The two reactive hydrogen groups in the primary amine react rapidly with the two epoxy functional groups in the epoxy resin by opening the epoxy ring and forming the secondary amine structure (inset diagram of Fig. 6b) [26], therefore improving the glass/polymer interface bonding. Further increase in silane concentration to 5 wt.% leads to the appearance of the protonated amine peak $-N^+$ (402.1 eV, Fig. 6c) which results from the reaction of the $NH_2-$ (or $-NH-$) groups in silane with the OH groups on the glass surface [21]. Thus, the protonated amino groups would have a preference towards the glass surface while the free amino groups protrude towards the epoxy side (inset diagram of Fig. 6c). Such chemical linkage at the interface between the glass and the epoxy accounts for the optimized glass coating/epoxy interfacial bonding in 5 wt.% silane modified sample.

[Fig. 6 here]

The surface chemistry of the silane modified glass also influences significantly the glass wettability (Fig. 6, d-f). For the as-prepared glass without silane, the contact angle between the epoxy and the glass surface is 45.6 º. By modifying the surface with 2 wt.% silane and 5 wt.% silane, the contact angle decreases to 33.7 º and 29.2 º, respectively. The lowest contact angle for 5 wt.% silane modified sample further confirms the strongest interfacial adhesion between glass-coated microwire and epoxy resin obtained in this sample.

3.2.3. Microwave properties of preset-damaged composites incorporating silane modified microwires arrays

To explore the relationship between stress transfer effectiveness enabled by the silane surface modification and the damage-sensing capabilities of the microwire composites, we investigated the microwave properties of the composites incorporating the silane modified microwires under preset damage. The damage was introduced by drilling a 1 mm diameter hole through the epoxy matrix along the thickness direction. The 1 mm-drilled hole was located at two different positions in the matrix: 2 mm (labeled as drilled-2mm) and 1 mm (labeled as drilled-1mm) away from the midpoint of microwire A (Fig. 1e). According to Fig. 7 a-d, the absorption and transmission spectra of the epoxy matrix exhibit basically the same type of dispersion with and without damage. This response is due to the almost transparency of the epoxy in the X-band and thus a small number of tiny holes would barely affect the effective permittivity of the matrix. Consequently, the embedded microwires will be directly responsible for the changes in the scattering spectra of the composites.

[Fig. 7 here]

For the composites with as-cast microwires, the transmission spectra show similar dispersion regardless of the presence and location of the damage (undrilled, drilled-2mm, drilled-1mm). However, the damage location can influence the absorption of the microwave in the composites. Compared to uncut free-standing arrays (absorption peak at 9.3 GHz, Fig. 3a), the absorption peak of undrilled as-cast wire composite samples blue-shifts to 10.4 GHz due to the compressive residual stresses generated by the different thermal expansion of microwire and epoxy during the curing process, inducing a non-axial magnetic anisotropy in the microwires. Moreover, the absorption dispersion of both the drilled-2 mm sample and the drilled-1 mm samples becomes broader and has smaller amplitude with respect to the undrilled as-cast wire composite samples. This is because during the development of the damage after drilling, the weak interfacial adhesion between glass-coated wire and the epoxy is destroyed, resulting in an inefficient epoxy-microwire stress transfer (Fig. 7e).

For the undrilled and drilled composites with 2 wt.% silane-modified microwires, the transmission spectra (Fig. 7a and b) show a similar trend with those incorporating as-cast microwires. However, the absorption curve illustrates a totally different trend (Fig. 7c and d). As previously discussed, by applying 2 wt.% silane, the glass-coated microwires are partially connected to the epoxy through free amino groups and thus compressive stresses are applied on the microwires by the epoxy resin inducing a strong uniaxial magnetic anisotropy. Such phenomena result in smaller absorption amplitude compared to the as-cast microwire composites. Additionally, when comparing the absorption response at higher frequency, no significant difference is found between the drilled-2mm 2 wt.% sample and the undrilled 2 wt.% sample. However, the absorption of the drilled-1mm sample is slightly higher than that of undrilled sample (Fig. 7d). In this case, the closer damage to microwires creates a significant stress concentration that leads to tiny cracks in the microwire and thus scattering of the microwave by the generated dipoles over the cracked surface.

For the composites incorporating 5 wt.% silane modified microwires, a large change in transmission is observed only for the drilled-1mm sample (Fig. 7b) while a change in absorption is evidenced for different damage locations (Fig. 7c and d). Thus the composites containing the 5 wt.% silane modified microwires show a great potential for damage monitoring enabled by a more efficient stress transfer across matrix-wire interface (Fig. 7e).

To better visualize the effect of silane treatment and damage location in the microwave properties of the microwire composites we calculated the rate of transmission change as shown in Fig. 7f and g. The transmission change for the epoxy matrix and all the drilled-2mm composite samples is almost the same with a change ratio less than 20%. This is mainly attributed to the metallic impurities introduced during drilling and/or measurement inaccuracy. Therefore, the sensing and damage detection capability through the microwave response of microwire composites is rather poor when the damage locates far from the wire inclusion. For the drilled-1mm samples, although the transmission change of as-cast microwire and 2 wt% silane modified microwires is not substantial, it dramatically changes up to 100% for 5wt% silane-modified microwires in the frequency range of 10-12 GHz. In this case, the microwave response of the composites overcomes the measurement uncertainties, providing reliable signals for damage detection. From Fig. 7g it is concluded that after being modified by 5wt% silane, the embedded microwire is able to sense the damage in close proximity to it, in contrast to the results discussed in Section 3.1 in which the microwires can only respond to damage when it locates right on them. Therefore, damage detection sensitivity of the microwire composites have increased thanks to the optimized interfacial conditions in the composite system. In fact, the 5% silane-modified glass surface serves as mediator in the stress transfer from the matrix to the microwire while preserving both the wire circumferential domain structure responsible for the sensing capabilities, and the structural integrity of the composite. Without such mediator the wire inclusion would have considerable different properties from the free-standing wire due to the wire/epoxy poor adhesion quality. In such situation, our basic understanding of the wire would be still valid in composite context which is highly desirable to predict composite properties. Overall the presence and optimization of the dual interface: epoxy/glass coating and glass coating/wire allowed us to achieve maximum damage detection sensitivity evaluated through microwave scattering. Additionally, it should be noted that the non-destructive monitoring method based on the microwave properties of glass-coated ferromagnetic microwires is limited by the microwire arrangement (wire amount and spacing) and damage location. To further refine the damage detection without adding more microwires to the array which could be unfavorable to structural integrity, we could modify the wire surface with e.g suitable nanoparticles or coatings capable of responding to external stimuli.

## 4   Conclusions

We have demonstrated the potential of developing a smart self-sensing composite enabled by the presence and optimization of dual interfaces in periodically arranged Co-based glass-coated microwire/epoxy matrix composites. In the first case where the damage was located directly on the wires, the microwave property was prescribed by the structural integrity expressed by the number of cracks, crack location and periodicity of the array. When the damage was introduced in the epoxy matrix, the microwire array modified by 5 wt.% silane could efficiently detect the local damage when it was located 1mm far from the nearest microwire in the array. Such enhanced damage detection capability was triggered in a favorable dual-interface context, which the glass cover served as a mediator between epoxy and wire core allowing the most efficient stress transfer and preserving both the wire superior magnetic properties and composite integrity. Our approach demonstrates the potential of regulating microwire composites sensitivity to external stimuli through surface modification which could fit specific requirements of structural health monitoring systems. It also opens up a new realm to achieve multifunctional composites with indiscriminate functionalities, i.e., developing composite fillers with dual or multiple interfaces to regulate the coupling and de-coupling effects as desired.

**Acknowledgements**

FXQ would like to thank the financial support of NSFC No. 51671171 and No. 5171101468; Basic Grants for Central Universities No. 2018QNA4009 and National Youth Thousand Talent Program' of China. All authors are grateful to Dr. Dmitriy Makhnovskiy for insightful discussions of experimental results.

**Figure Captions:**

Fig. 1. (a) The schematic of paper frame used to place the free-standing wires inside the rectangular waveguide. (b) The pre-set cracks setup for the free-standing microwire array. The capital letter represents the microwire with crack on it. The number following the letter represents the distance (mm) between the microwire top and the crack. (c) Surface modification of microwire by using KH550 silane. (d) The microwire is embedded in the epoxy, with 1 mm effective pull-out length. (e) Holes introduced into the matrix to imitate the situation in which damage locates in the composite matrix near the microwire.

Fig. 2. (a)The absorption spectra and (b) the reflection spectra of free-standing microwire

Fig, 3. (a)The absorption spectra and (b) the reflection spectra of free-standing microwire arrays with two crack.

Fig. 4. Pull-out result for surface-modified microwires embedded by epoxy.

Fig. 5. Diagrams and Scanning electron microscopy (SEM) images for pulled-out microwires (a) As-cast and modified by (b) 2 wt.% and (c) 5 wt.% silane.

Fig. 6. XPS spectrum for (a) as-prepared glass and glass modified by (b) 2 wt.% and (c) 5 wt.% KH550 silane. Contact angle of (d) as-prepared glass and glass modified by (e) 2 wt.% and (f) 5 wt.% silane to epoxy.

Fig. 7. Transmission for (a) drilled-2 mm composites samples (b) drilled-1 mm composites samples. Absorption for (c) drilled-2 mm composites samples and (d) drilled-1 mm composites samples. (e) Stress transfer diagrams for as cast, 2 wt% KH550 and 5 wt% KH550 samples. Rate of transmission change for (f) drilled-2 mm composites samples and (g) drilled-1 mm composites samples

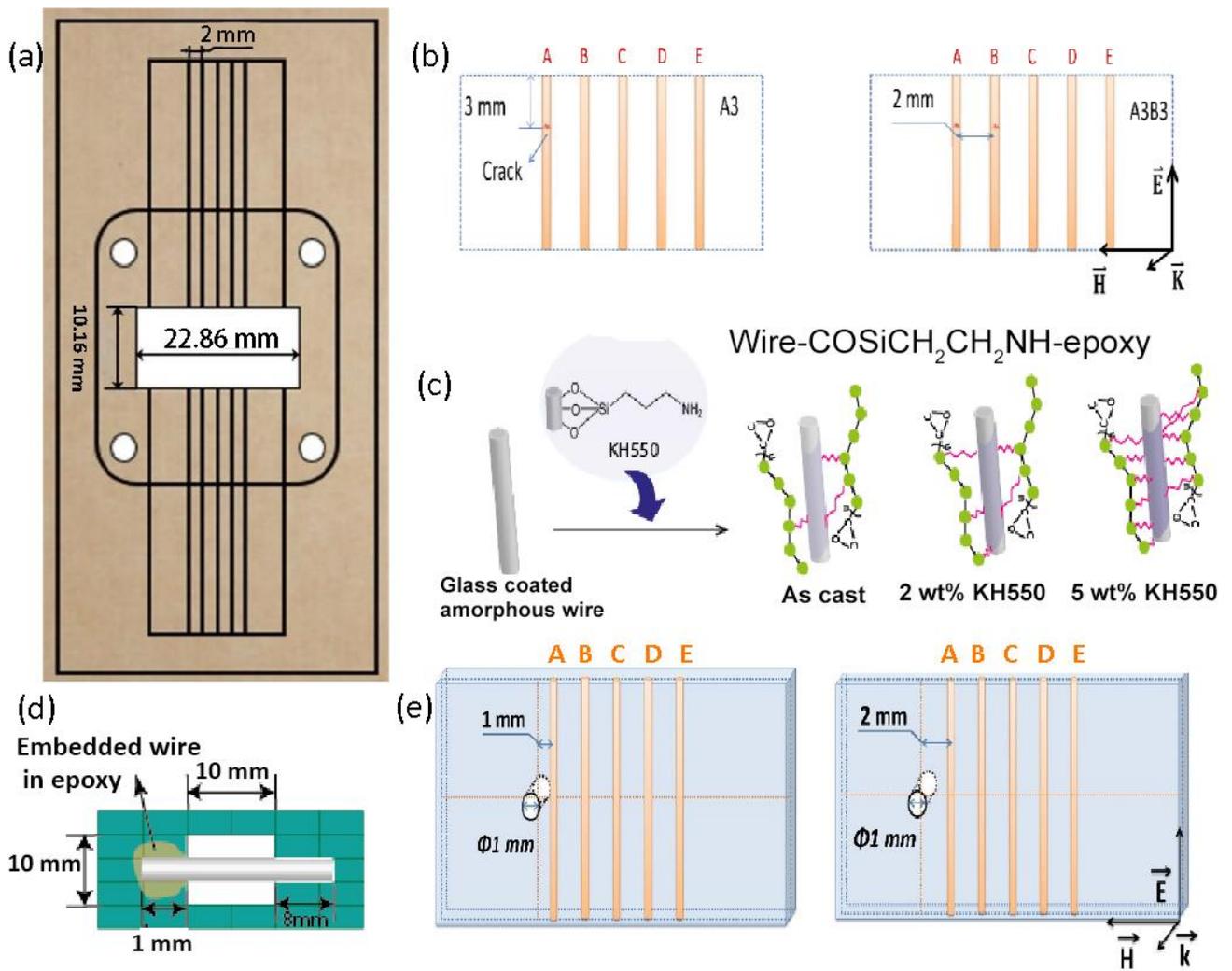

Fig. 1

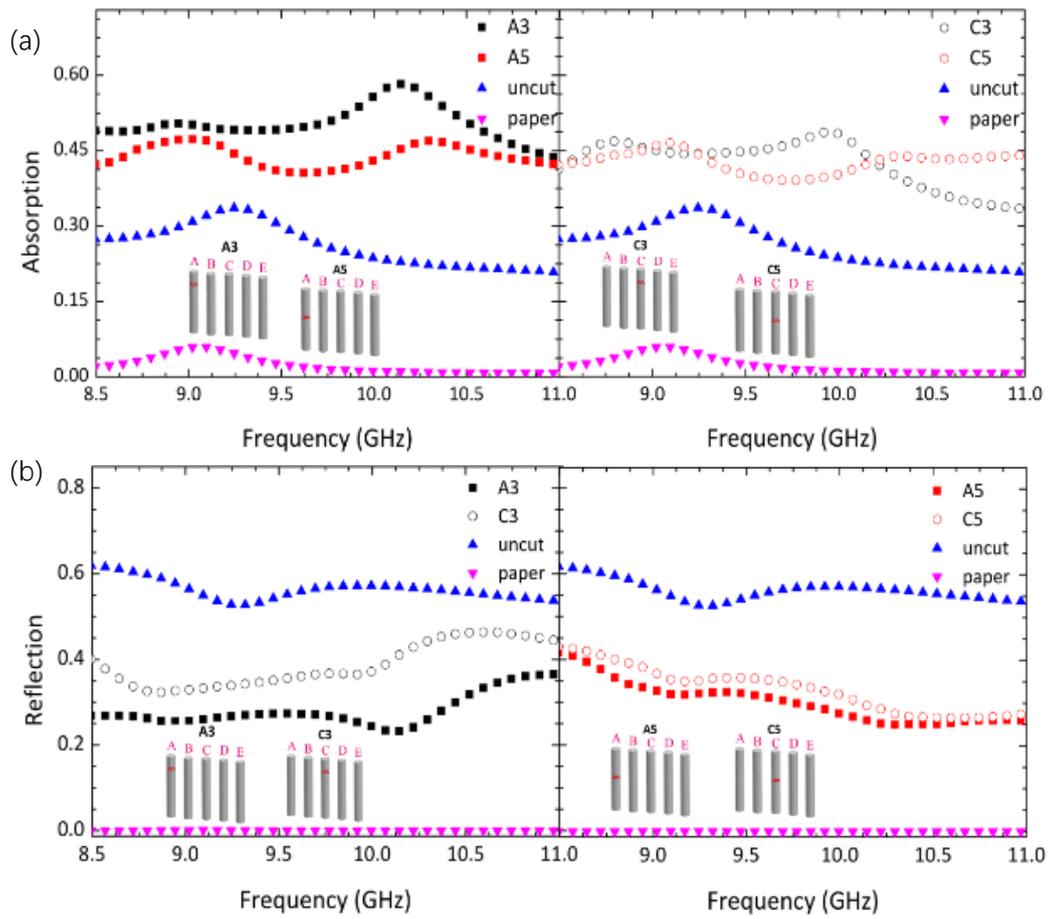

**Fig. 2**

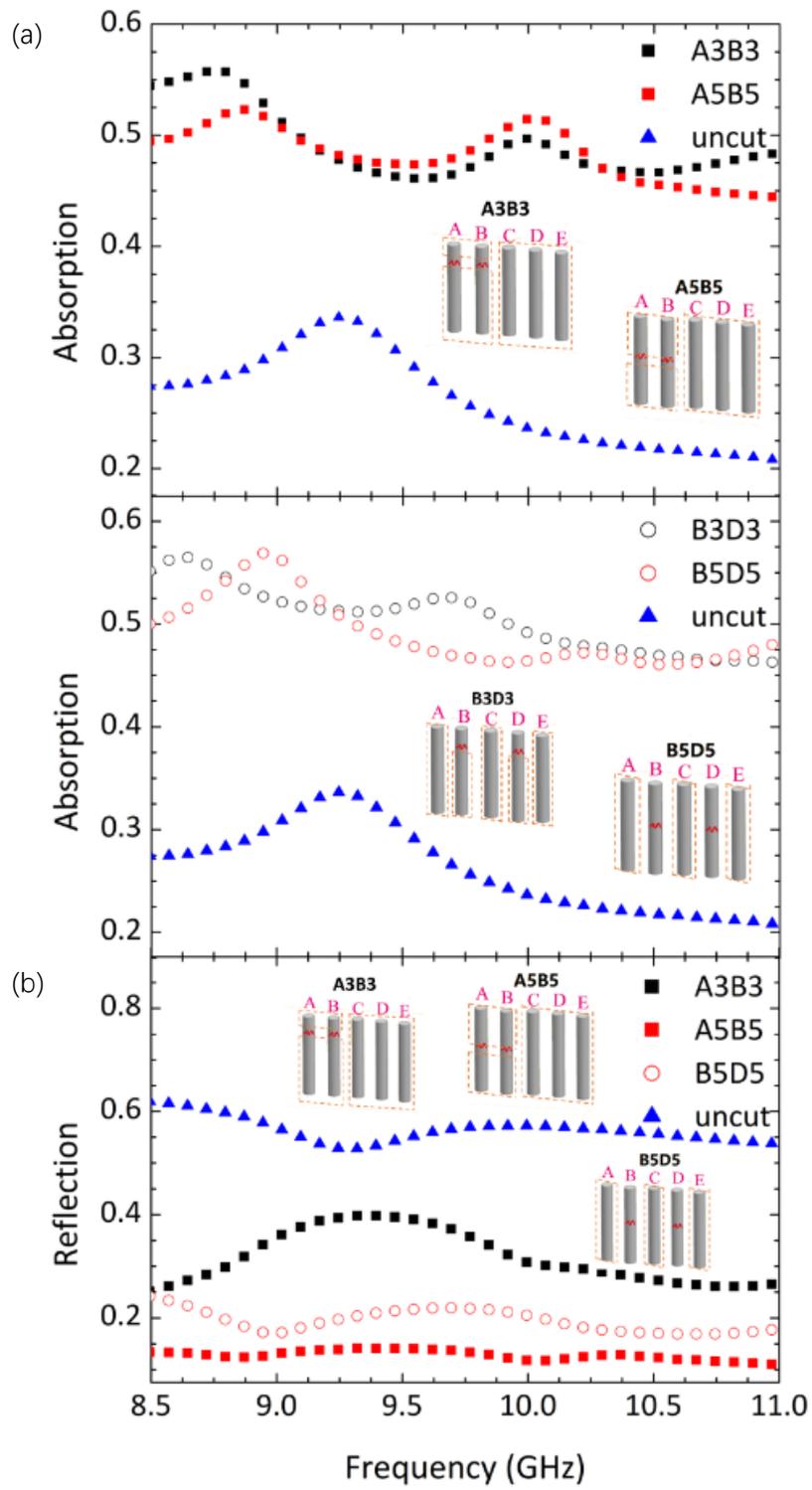

Fig. 3

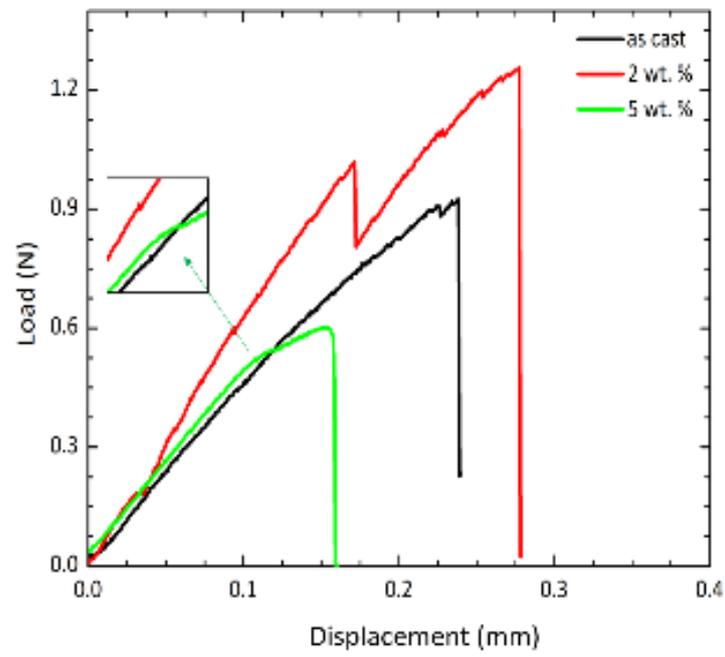

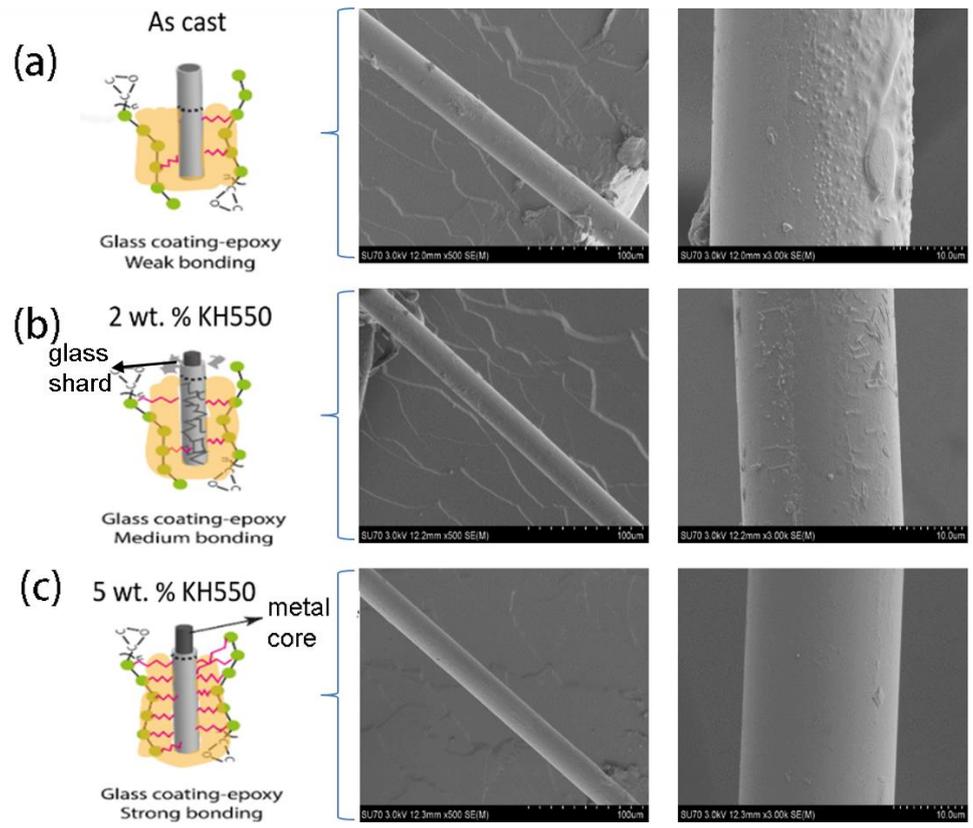

**Fig. 5**

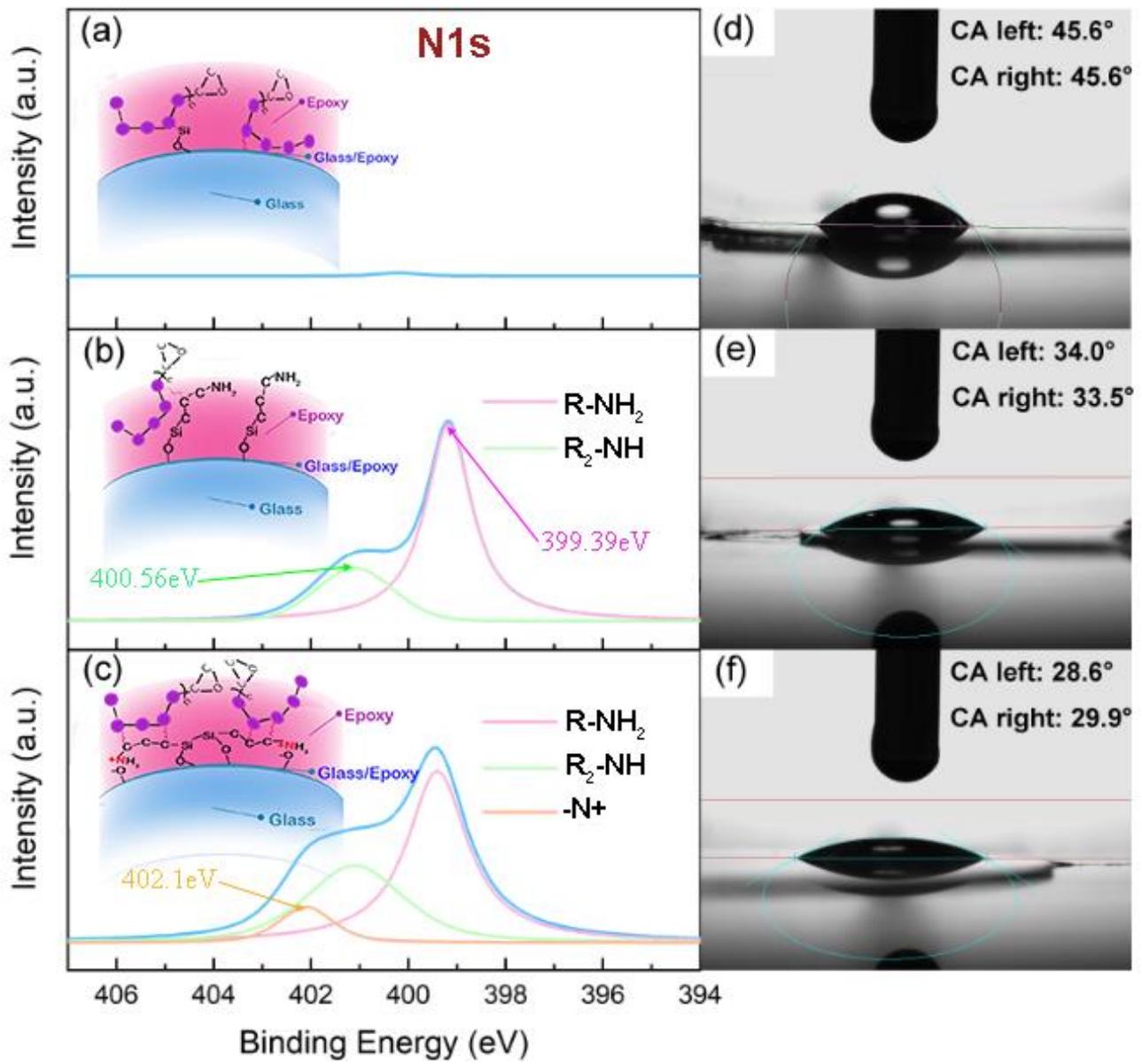

**Fig. 6**

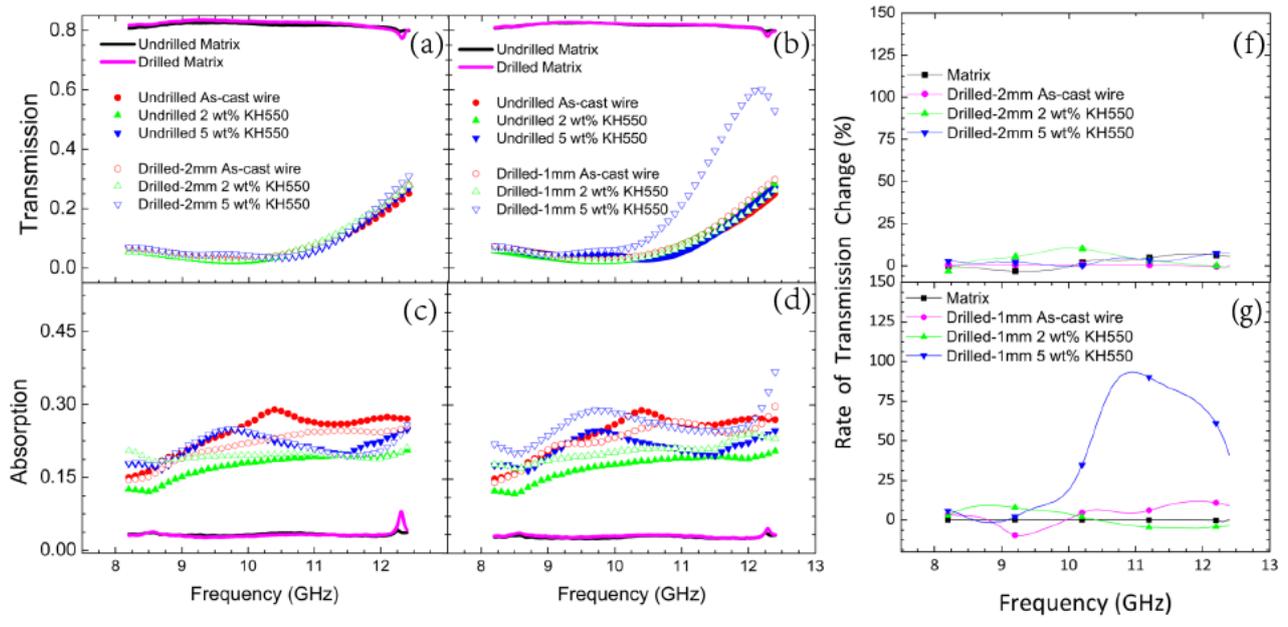

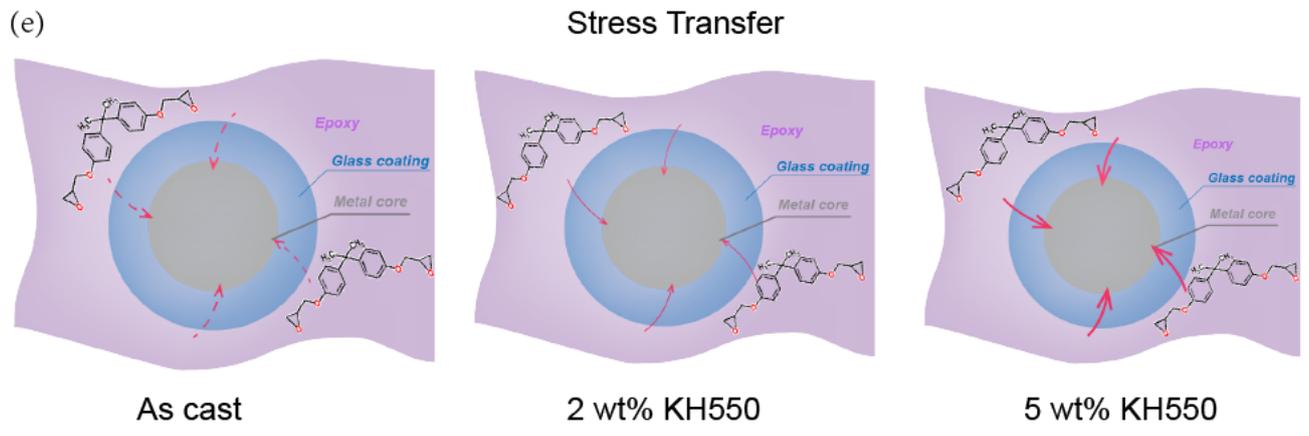

**Stress Transfer**

As cast          2 wt% KH550          5 wt% KH550

**Fig. 7**

Table 1 Types of cracks preset on the microwire arrays.

| Number of cracks | Type |
| --- | --- |
| 1 | A3; C3; A5; C5 |
| 2 | A3B3; A5B5; B3D3; B5D5 |

Table 2 The concentration of silane solution is 2 wt.% and 5 wt.%, respectively. The microwires are immersed in silane solution in an ultrasonic bath for 20 minutes.

| Mass Fraction | Ethanol (g) | KH550 (g) |
| --- | --- | --- |
| 2 wt.% | 49.0 | 1.0 |
| 5 wt.% | 47.5 | 2.5 |